\documentclass[12pt,preprint2]{aastex6}

\usepackage{amsmath}
\usepackage{natbib}
\newcommand{\oau}{\rm{[\ion{O}{3}]$\lambda$4363}}
\newcommand{\oin}{\rm{\ion{O}{3}]$\lambda\lambda$1661,1666}}
\newcommand{\nii}{\rm{[\ion{N}{2}]$\lambda$6584}}
\newcommand{\oiii}{\rm{[\ion{O}{3}]$\lambda$5007}}
\newcommand{\oiiid}{\rm{[\ion{O}{3}]$\lambda\lambda$4959,5007}}
\newcommand{\oii}{\rm{[\ion{O}{2}]$\lambda$3727}}
\newcommand{\oiid}{\rm{[\ion{O}{2}]$\lambda\lambda$3726,3729}}
\newcommand{\ha}{\rm H{$\alpha$}}
\newcommand{\hb}{\rm H{$\beta$}}

\newcommand{\siid}{\rm{[\ion{S}{2}]$\lambda\lambda$6717,6731}}
\newcommand{\niid}{\rm{[\ion{N}{2}]$\lambda\lambda$6548,6584}}
\newcommand{\oi}{\rm[\ion{O}{1}]$\lambda$6300}
\newcommand{\neiii}{\rm[\ion{Ne}{3}]$\lambda$3869}
\newcommand{\gmass}{\log(M_*/M_{\sun})}

\slugcomment{}        

\shorttitle{}
\shortauthors{F. Bian et al.}

\begin{document}


\title{``Direct" Gas-Phase Metallicity in Local Analogs of High-Redshift Galaxies: Empirical Metallicity Calibrations for High-Redshift Star-Forming Galaxies}


\author{Fuyan Bian\altaffilmark{1,2}, Lisa J. Kewley\altaffilmark{1,3}, Michael A. Dopita\altaffilmark{1,3}}
\altaffiltext{1}{Research School of Astronomy and Astrophysics, Australian National University, Canberra, ACT 2611, Australia}
\altaffiltext{2}{European Southern Observatory, Alonso de C\'ordova 3107, Casilla 19001, Vitacura, Santiago 19,
Chile}
\altaffiltext{3}{ARC Centre of Excellence for All Sky Astrophysics in 3 Dimensions (ASTRO 3D)}



\begin{abstract}
We study the direct gas-phase oxygen abundance using the well-detected auroral line {\oau} in the stacked spectra of a sample of local analogs of high-redshift galaxies. These local analogs share the same location as $z\sim2$ star-forming galaxies on the {\oiii/\hb} versus {\nii/\ha} Baldwin-Phillips-Terlevich  diagram. This type of analog has the same ionized interstellar medium (ISM) properties as high-redshift galaxies. We establish empirical metallicity calibrations between the direct gas-phase oxygen abundances ($7.8<12+\log(\rm{O/H})<8.4$) and the N2 ($\log$(\nii/H$\alpha$))/O3N2 ($\log$((\oiiid/\hb)/(\nii/\ha))) indices in our local analogs. We find significant systematic offsets between the metallicity calibrations for our local analogs of high-redshift galaxies and those derived from the local \ion{H}{2} regions and a sample of local reference galaxies selected from the Sloan Digital Sky Survey (SDSS). The N2 and O3N2 metallicities will be underestimated by 0.05-0.1~dex relative to our calibration, if one simply applies the local metallicity calibration in previous studies to high-redshift galaxies. Local metallicity calibrations also cause discrepancies of metallicity measurements in high-redshift galaxies using the N2 and O3N2 indicators. In contrast, our new calibrations produce consistent metallicities between these two indicators.  We also derive metallicity calibrations for R23 ($\log$((\oiiid+\oiid)/\hb)), O32($\log$(\oiiid/\oiid)),  $\log($\oiii/\hb), and $\log$({\neiii}/{\oii}) indices in our local analogs, which show significant offset compared to those in the SDSS reference galaxies. By comparing with MAPPINGS photoionization models, the different empirical metallicity calibration relations in the local analogs and the SDSS reference galaxies can be shown to be primarily due to the change of ionized ISM conditions.  Assuming that temperature structure variations are minimal and ISM conditions do not change dramatically from $z\sim2$ to $z\sim5$, these empirical calibrations can be used to measure relative metallicities in galaxies with the redshifts up to $z\sim5.0$ in ground-based observations.

\end{abstract}



\section{Introduction}
Studies of galaxy chemical abundances across cosmic history provide insight into the key physical processes governing the formation and evolution of galaxies \citep[e.g.,][]{Finlator:2008fj,Dave:2012aa,Lilly:2013aa,Lu:2015aa,Ma:2016aa}. A relationship between galaxy stellar mass and gas-phase oxygen abundance has been well established in the local universe \citep{Tremonti:2004aa,Savaglio:2005aa,Kewley:2008aa,Andrews:2013aa}. Galaxies with higher stellar masses tend to have higher metallicities than galaxies with lower stellar masses. An evolution of the mass-metallicity relation has been found in galaxies up to a redshift of $z\sim3.5$ \citep[][]{Erb:2006rt,Maiolino:2008lr,Zahid:2013aa,Zahid:2014aa,Maier:2014aa,Steidel:2014aa,Sanders:2015aa,Guo:2016aa,Ly:2016ab,Onodera:2016aa}. High-redshift star-forming galaxies have lower metallicities for a given stellar mass compared to galaxies at lower redshift. 

There are three ways to measure the gas-phase oxygen abundances:
\begin{enumerate} 
\item Theoretical metallicity calibrations, which combine stellar synthesis models \citep[e.g., Starburst99,][]{Leitherer:1999aa} and photoionization models, including MAPPINGS \citep{Sutherland:1993aa,Dopita:2013aa} and CLOUDY \citep{Ferland:2013aa}, to predict the metallicity-sensitive emission-line ratios based on input metallicities \citep[e.g.,][]{McGaugh:1991aa,Kewley:2002fk,Tremonti:2004aa,Dopita:2016aa}. 
\item The $``T_e"$ or ``direct" method, which measures electron temperature using the ratio of the {\oau} auroral emission line to the {\oiii} emission line. This method requires one to correct for unseen stages of ionization and assume a uniform electron temperature throughout the gas. The metallicity is then estimated based on the electron temperature \citep[e.g.,][]{Pagel:1992aa,Izotov:2006ab}. 
\item Empirical metallicity calibrations, which establish relations between the $T_e$ metallicity and the strong line ratios to overcome the difficulty of detecting the weak {\oau} auroral emission line \citep[e.g.,][]{Pettini:2004qe,Pilyugin:2005aa,Nagao:2006aa,Pilyugin:2010aa,Pilyugin:2012aa,Perez-Montero:2009aa,Marino:2013aa, Brown:2016aa,Cowie:2016aa, Curti:2017aa}. For example, \citet[PP04 hereafter]{Pettini:2004qe} fit the relationship between the direct $T_e$-based metallicities\footnote{PP04 also used photoionization models to measure the metallicities for a small fraction of high-metallicity \ion{H}{2} regions.} and strong emission-line ratios, including the $N2 = \log$(\nii/\ha) and $O3N2 =\log$[(\oiii/\hb)/(\nii/\ha)] indices in a sample of \ion{H}{2} regions in nearby star-forming galaxies. These empirical calibrations based on local \ion{H}{2} regions have been widely used to estimate the metallicities in high-redshift galaxies \citep[e.g.,][]{Erb:2006rt,Hainline:2009aa,Bian:2010vn,Steidel:2014aa,Sanders:2015aa}, though it is unclear whether these empirical calibrations are still feasible for high-redshift star-forming galaxies.  
\end{enumerate}

\begin{figure*}[ht!]
\begin{center}
\includegraphics[width=2\columnwidth]{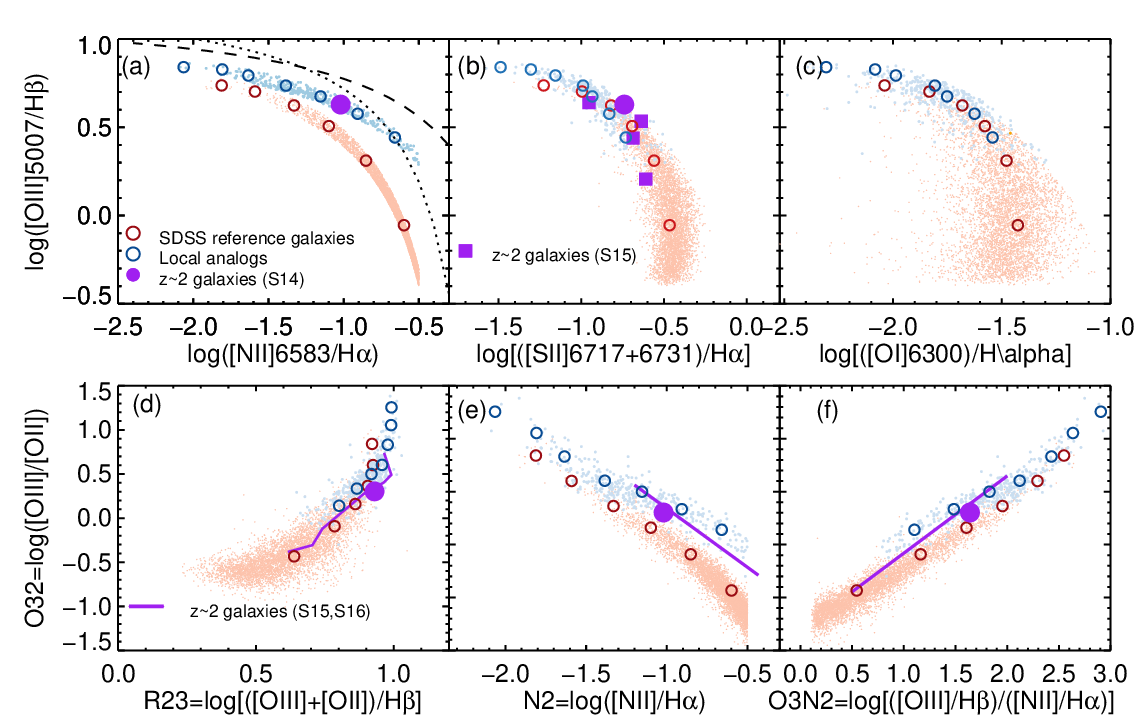}
\caption{Diagnostic diagrams of local analogs of high-redshift galaxies and SDSS reference galaxies. The small blue and red points represent the individual local analogs and SDSS reference galaxies, respectively, and the large open blue and red points represent the stacked spectra of the local analogs and SDSS reference galaxies, respectively. The large purple filled circle represents the stacked spectrum of a sample of $z\sim2$ UV-selected galaxies adopted from \citet[S14]{Steidel:2014aa}.  The purple squares represent the median distribution of a sample of $z\sim2$ mass-selected galaxies adopted from \citet[S15]{Shapley:2015aa}, and the purple solid lines represent the median distribution of a sample of $z\sim2$ mass-selected galaxies adopted from \citet[S15]{Shapley:2015aa} and \citet[S16]{Sanders:2016aa}. The dotted and dashed lines in panel (a) represent the empirical \citep{Kauffmann:2003ij} and theoretical \citep{Kewley:2001aa} separations of star-forming galaxies
and AGNs, respectively.
 \label{BPT}}
\end{center}
\end{figure*}

Photoionization models show that the {\nii/\ha} and {\oiii/\hb} ratios depend not only on the metallicity but also on other effects, including the ionization parameter, the electron density, the nitrogen-to-oxygen ratio (N/O), and the spectral shape of the radiation field \citep[e.g.,][]{Kewley:2002fk,Kewley:2013ab,Kewley:2013aa,Cullen:2016aa,Hirschmann:2017aa}. If one or a combination of these factors changes with cosmic time, metallicity calibrations derived from local galaxies will not be applicable to high-redshift galaxies \citep[e.g.,][]{Bian:2017aa}. Changes in the physical conditions of the ionized ISM condition and/or the radiation field can also shift the location of high-redshift galaxies in the {\oiii/\hb} versus {\nii/\ha} ``Baldwin-Phillips-Terlevich'' \citep[BPT,][]{Baldwin:1981rr,Veilleux:1987aa} diagram and in other commonly used diagnostic diagrams \citep[e.g., Figure~\ref{BPT},][]{Kewley:2013ab,Kewley:2013aa,Kewley:2016aa,Steidel:2014aa,Shapley:2015aa,Sanders:2016aa,Bian:2016aa}. Using these diagrams, studies have shown that the ionization parameter, electron density \citep[e.g.,][]{Brinchmann:2008ab,Liu:2008aa,Kewley:2013ab,Kewley:2013aa,Nakajima:2014aa,Shirazi:2014ab,Sanders:2016aa,Kaasinen:2017aa}, hardness of the stellar radiation fields \citep[e.g.,][]{Steidel:2014aa,Strom:2017aa, Strom:2017ab}, and/or the relation between N/O and metallicity \citep[e.g.,][]{Masters:2014aa,Shapley:2015aa} might change across cosmic time. These potential changes raise a crucial question: are the empirical calibrations based on the local \ion{H}{2} regions still valid for high-redshift galaxies?

There are two methods by which to resolve this issue:
\begin{enumerate}
\item Implementing the effects of changing ISM conditions, radiation field, and N/O within photoionization models to obtain metallicity diagnostics for high-redshift galaxies that explicitly take these changes into account. Unfortunately for this approach, there exist degeneracies when using the BPT and other optical diagnostic diagrams to study how the above effects vary with redshift. It is still under debate which one or a combination of the above effects change with redshift and by how much.  
\item Establishing the empirical relation between direct $``T_e"$ metallicity and the strong emission line in high-redshift galaxies. The ``direct" gas-phase oxygen abundance has been well measured up to the redshift of $z\simeq1.0$ \citep[e.g.,][]{Ly:2014aa,Ly:2015aa,Ly:2016aa,Ly:2016ab,Jones:2015ab}. However, the {\oau} emission line is rarely detected in high-redshift galaxies at $z\sim2$ \citep[e.g.,][]{Yuan:2009aa,Rigby:2011aa,Christensen:2012ys,Sanders:2016ab}. Alternatively, the {\oin} inter-combination emission line also can be used to measure the direct metallicity, however, the sample size is still small \citep[e.g.,][]{James:2014aa,Steidel:2014aa,Steidel:2016aa,Kojima:2016aa,Vanzella:2016aa}.
\end{enumerate}

In this paper, we use a sample of local analogs of high-redshift galaxies selected from the Sloan Digital Sky Survey (SDSS) to study the empirical calibration at $z\sim2$. These local analogs share the same region on the BPT diagram with $z\sim2$ star-forming galaxies from \citet{Steidel:2014aa}. \citet{Bian:2016aa} demonstrated that the galaxies with this BPT location selection have the same ionized ISM properties of $z\sim2$ galaxies. Here, we use the stacked spectrum of both our analogs and the SDSS reference galaxies to detect the weak {\oau} line and measure the direct $T_e$ metallicity, which enables us to compare how the empirical metallicity calibration changes with redshift. This paper is organized as follows. In Section~\ref{sample}, we select an SDSS reference galaxy sample and a sample of local analogs of high-redshift galaxies. In Section~\ref{analysis}, we measure the direct $T_e$ oxygen abundance in the stacked spectra of the SDSS reference galaxy sample and the local analog galaxy sample in different $N2$ bins. In Section~\ref{results}, we establish the relation between the direct $T_e$ oxygen abundance and the metallicity diagnostic line ratios to study how the changes in the ISM conditions affect these empirical metallicity calibrations. In Section~\ref{discussion}, we compare our new empirical metallicity calibrations with other empirical calibrations and photoionization models. In Section~\ref{conclusion}, we summarize the main results of this paper.

\section{Sample Selection}\label{sample}
In this section, we select a sample of local reference galaxies and local analogs of high-redshift galaxies from the SDSS \citep{York:2000aa} MPA-JHU value-added catalog for SDSS Data Release 7 \citep[][]{Abazajian:2009aa}. The MPA-JHU catalog includes 819,333 galaxies, which are the part of the
SDSS ``main" galaxy sample \citep{Strauss:2002aa}. The MPA-JHU catalog contains information on the fluxes, equivalent widths, and line widths of the optical emission lines, including [\ion{O}{2}]$\lambda3727$, H$\beta$, [\ion{O}{3}]$\lambda\lambda4959, 5007$, H$\alpha$, [NII]$\lambda6583$, and [\ion{S}{2}]$\lambda\lambda6717,6731$, stellar masses \citep[$M_*$,][]{Kauffmann:2003aa}, star formation rates \citep[SFRs,][]{Brinchmann:2004aa}, and specific SFRs \citep[SFR/$M_{*}$, sSFRs,][]{Brinchmann:2004aa}. We first select galaxies that were classified as star-forming or starburst in the MPA-JHU catalog. Then we require the signal-to-noise ratios (S/Ns) of [\ion{O}{2}]$\lambda$3727, H$\beta$, [\ion{O}{3}]$\lambda\lambda4959, 5007$, H$\alpha$ and [\ion{N}{2}]$\lambda$6583 emission lines greater than 10 in the MPA-JHU catalog. At last, we select the local reference galaxies and local analogs based on their locations on the BPT diagram.

\subsection{SDSS Reference Galaxy Sample}
We select local reference galaxies located in the $\pm0.05$~dex region of the local star-forming sequence defined by the equation 3 in \citet{Kewley:2013ab} on the BPT diagram as follows (red small data points in Figure~\ref{BPT}(a)):

\begin{equation}
\log({\rm[OIII]/H\beta)})>\frac{0.61}{\log({\rm[NII]/H\alpha)})-0.08}+1.05 \label{lz_lower},
\end{equation}
\begin{equation}
\log({\rm[OIII]/H\beta)})<\frac{0.61}{\log({\rm[NII]/H\alpha)}-0.08}+1.15 \label{lz_upper},
\end{equation} 
and 
\begin{equation}
\log({\rm[NII]/H\alpha)} < -0.5\label{lz_right}
\end{equation}

The following [\ion{S}{2}]$\lambda\lambda6717,6731$ and {\oi} diagnostics are applied to further remove galaxies with AGN/shock contamination to the emission-line flux \citep{Kewley:2006aa}
\begin{equation}
\log({\rm[OIII]/H\beta}) < \frac{0.72}{\log({\rm [SII]/H\alpha})- 0.32} + 1.30 \label{lz_s2}
\end{equation}
\begin{equation}
\log({\rm[OIII]/H\beta}) < \frac{0.73}{\log({\rm [OI]/H\alpha})- 0.59} + 1.33 \label{lz_o1}
\end{equation}

A total of 22428 unique SDSS galaxies are selected. In general, these galaxies represent low-redshift star-forming galaxies, which are located on the local BPT star-forming sequence. 

\subsection{Local Analogs of High-redshift Galaxies}\label{analog}
The sample of local analogs of high-redshift galaxies has been selected from in the $\pm0.04$~dex region of the $z\sim2.3$ star-forming sequence defined by equation~9 in \citet{Steidel:2014aa} on the BPT diagram (blue small data points in Figure~\ref{BPT}(a)):
\begin{equation}
\log({\rm[OIII]/H\beta)})>\frac{0.67}{\log({\rm[NII]/H\alpha)})-0.33}+1.09 \label{hz_lower},
\end{equation}
\begin{equation}
\log({\rm[OIII]/H\beta)})<\frac{0.67}{\log({\rm[NII]/H\alpha)})-0.33}+1.17 \label{hz_upper},
\end{equation} 
and
\begin{equation}
\log({\rm[NII]/H\alpha)}) < -0.5\label{hz_right}
\end{equation}

All the galaxies selected by equations of \ref{hz_lower} - \ref{hz_right} are classified as galaxies by the \citet{Kewley:2001aa} criterion (Figure~\ref{BPT}(a)), and all the galaxies with $\log$({\nii}/{\ha}$)<-0.75$ are classified as galaxies by the \citet{Kauffmann:2003ij} criterion. We further apply the [\ion{S}{2}] and [\ion{O}{1}] BPT diagnostic to reduce the AGN/shock contaminants. Studies have shown that $z\sim2$ star-forming galaxies exhibit a higher [\ion{O}{3}]/H$\beta$ ratio for a given [\ion{S}{2}]/{\ha} ratio \citep[e.g.,][]{Strom:2017aa}. Thus, we shift
the \citet{Kewley:2006aa} criteria on the [\ion{O}{3}]/H$\beta$ ratio by +0.05~dex for high-redshift galaxies. 

\begin{equation}
\log({\rm[OIII]/H\beta}) < \frac{0.72}{\log({\rm [SII]/H\alpha})- 0.32} + 1.35 \label{lz_s2}
\end{equation}
\begin{equation}
\log({\rm[OIII]/H\beta}) < \frac{0.73}{\log({\rm [OI]/H\alpha})- 0.59} + 1.38 \label{lz_o1}
\end{equation}

A total of 443 galaxies are selected based on the above selection criteria. Equations~\ref{lz_s2} and \ref{lz_o1} remove about 20\%, and these two criteria do not affect the distribution of the local analogs on the optical diagnostics diagrams. We cross-correlate the positions of the analogs of high-redshift galaxies with the {\it ROSAT} ALL-Sky Survey Faint Source Catalog \citep{Voges:2000aa} and find that none of the analogs are detected at X-ray wavelengths. We further check the local analogs with high-redshift galaxies with high full width half maximum of the Balmer emission lines to remove potential AGNs hosting low-mass supermassive black holes \citep[$10^6~M_{\sun}$ with $\rm FWHM >600$ km~s$^{-1}$,][]{Greene:2004aa}. We find only one galaxy with $\rm FWHM >400$ km~s$^{-1}$ among our local analogs, and the galaxy is removed from our further analysis. We also visually inspect the image of the local analogs and their fiber positions to remove spurious galaxies being targeted due to poor photometric deblending. The properties of these local analogs can be summarized as follows. The median stellar mass is $\gmass=8.8\substack{+0.06 \\ -0.02}$. The median SFR and sSFR are $3.9\substack{+0.7 \\ -0.2}$~$M_{\sun}$~yr$^{-1}$ and $10.0\substack{+1.0 \\ -0.5}$~Gyr$^{-1}$, respectively. The sSFR of the local analogs is comparable to that in $z\sim2$ star-forming galaxies with similar stellar mass \citep[e.g.,][]{Rodighiero:2011fk}. Furthermore, these analogs closely resemble the ISM conditions in high-redshift galaxies, including high ionization parameters ($\log q\simeq7.9$~cm$^{-1}$) and high electron densities ($n_e\simeq120$~cm$^{-3}$). The median ionization parameters and electron densities in these analogs are also comparable to those in the $z\sim2-3$ galaxies \citep[e.g.,][]{Nakajima:2014aa,Sanders:2016aa}. We refer readers to \citet{Bian:2016aa,Bian:2017aa} for more details.

\begin{table*}
\centering 
\caption{The Relative Line Flux from the Stacked Spectra of the SDSS Reference Galaxies\label{tab:SDSS_flux}}
\begin{tabular}{lcccccccccccccccccc}
\hline
line&N2$_{-0.75}^{-0.50}$&N2$_{-1.00}^{-0.75}$&N2$_{-1.25}^{-1.00}$&N2$_{-1.50}^{-1.25}$&N2$_{-1.75}^{-1.50}$&N2$_{-2.00}^{-1.75}$&N2$_{-2.25}^{-2.00}$\cr
\hline
\hline
[\ion{O}{2}]$\lambda3727$&$  3.301\pm  0.038$&$  3.513\pm  0.023$&$  3.088\pm  0.014$&$  2.523\pm  0.011$&$  1.750\pm  0.009$&$  1.071\pm  0.007$&$  0.566\pm  0.011$\cr
[\ion{Ne}{3}]$\lambda3869$&$  0.125\pm  0.008$&$  0.198\pm  0.007$&$  0.287\pm  0.007$&$  0.368\pm  0.009$&$  0.427\pm  0.008$&$  0.451\pm  0.007$&$  0.512\pm  0.010$\cr
[\ion{O}{3}]$\lambda4363$&$  0.006\pm  0.004$&$  0.015\pm  0.002$&$  0.027\pm  0.001$&$  0.045\pm  0.001$&$  0.077\pm  0.001$&$  0.116\pm  0.001$&$  0.156\pm  0.002$\cr
H$\beta$&$  1.000\pm  0.073$&$  1.000\pm  0.073$&$  1.000\pm  0.080$&$  1.000\pm  0.102$&$  1.000\pm  0.235$&$  1.000\pm  0.383$&$  1.000\pm  0.865$\cr
[\ion{O}{3}]$\lambda4959$&$  0.290\pm  0.003$&$  0.679\pm  0.003$&$  1.069\pm  0.003$&$  1.407\pm  0.003$&$  1.672\pm  0.006$&$  1.824\pm  0.008$&$  2.198\pm  0.011$\cr
[\ion{O}{3}]$\lambda5007$&$  0.878\pm  0.007$&$  2.038\pm  0.009$&$  3.199\pm  0.009$&$  4.190\pm  0.010$&$  5.022\pm  0.017$&$  5.453\pm  0.022$&$  6.660\pm  0.033$\cr
[\ion{O}{1}]$\lambda6100$&$  0.107\pm  0.002$&$  0.095\pm  0.001$&$  0.076\pm  0.000$&$  0.060\pm  0.000$&$  0.042\pm  0.001$&$  0.026\pm  0.001$&$  0.014\pm  0.001$\cr
[\ion{N}{2}]$\lambda6548$&$  0.233\pm  0.002$&$  0.130\pm  0.001$&$  0.073\pm  0.000$&$  0.043\pm  0.001$&$  0.023\pm  0.001$&$  0.014\pm  0.001$&$  0.010\pm  0.001$\cr
H$\alpha$&$  2.860\pm  0.016$&$  2.860\pm  0.009$&$  2.860\pm  0.006$&$  2.860\pm  0.005$&$  2.860\pm  0.007$&$  2.860\pm  0.009$&$  2.860\pm  0.010$\cr
[\ion{N}{2}]$\lambda6584$&$  0.720\pm  0.004$&$  0.402\pm  0.001$&$  0.228\pm  0.001$&$  0.134\pm  0.001$&$  0.074\pm  0.001$&$  0.044\pm  0.001$&$  0.026\pm  0.001$\cr
[\ion{S}{2}]$\lambda6717$&$  0.529\pm  0.001$&$  0.421\pm  0.000$&$  0.311\pm  0.000$&$  0.233\pm  0.000$&$  0.155\pm  0.001$&$  0.094\pm  0.001$&$  0.051\pm  0.001$\cr
[\ion{S}{2}]$\lambda6731$&$  0.372\pm  0.001$&$  0.299\pm  0.000$&$  0.222\pm  0.000$&$  0.166\pm  0.000$&$  0.111\pm  0.001$&$  0.069\pm  0.001$&$  0.040\pm  0.001$\cr
used&n&n&y&y&y&y&n\cr
$T_e$ Metallicity&  -&  -&$  8.26\pm  0.02$&$  8.18\pm  0.009$&$  8.00\pm  0.008$&$  7.80\pm  0.006$&  -\cr

\hline

 \end{tabular}
\end{table*}

\begin{table*}
\centering
\caption{The Relative Line Flux in the Spectra of the Local Analogs of High-redshift Galaxies.\label{tab:analog_flux}}
\begin{tabular}{lcccccccccccccccccc}

\hline
line&N$_{-0.75}^{-0.50}$&N$_{-1.00}^{-0.75}$&N$_{-1.25}^{-1.00}$&N$_{-1.50}^{-1.25}$&N$_{-1.75}^{-1.50}$&N$_{-2.00}^{-1.75}$&N$_{-2.25}^{-2.00}$\cr
\hline
\hline
[\ion{O}{2}]$\lambda3727$&$  2.757\pm  0.030$&$  2.404\pm  0.023$&$  2.071\pm  0.011$&$  1.881\pm  0.008$&$  1.264\pm  0.005$&$  0.819\pm  0.012$&$  0.532\pm  0.007$\cr
[\ion{Ne}{3}]$\lambda3869$&$  0.245\pm  0.010$&$  0.313\pm  0.008$&$  0.400\pm  0.007$&$  0.464\pm  0.008$&$  0.514\pm  0.007$&$  0.514\pm  0.007$&$  0.537\pm  0.010$\cr
[\ion{O}{3}]$\lambda4363$&$  0.017\pm  0.003$&$  0.024\pm  0.001$&$  0.043\pm  0.001$&$  0.063\pm  0.001$&$  0.095\pm  0.001$&$  0.132\pm  0.001$&$  0.167\pm  0.002$\cr
H$\beta$&$  1.000\pm  0.007$&$  1.000\pm  0.007$&$  1.000\pm  0.004$&$  1.000\pm  0.003$&$  1.000\pm  0.003$&$  1.000\pm  0.005$&$  1.000\pm  0.007$\cr
[\ion{O}{3}]$\lambda4959$&$  0.899\pm  0.007$&$  1.244\pm  0.009$&$  1.563\pm  0.006$&$  1.797\pm  0.005$&$  2.059\pm  0.006$&$  2.251\pm  0.011$&$  2.348\pm  0.016$\cr
[\ion{O}{3}]$\lambda5007$&$  2.761\pm  0.019$&$  3.761\pm  0.025$&$  4.713\pm  0.018$&$  5.427\pm  0.015$&$  6.196\pm  0.016$&$  6.705\pm  0.031$&$  6.912\pm  0.047$\cr
[\ion{O}{1}]$\lambda6100$&$  0.082\pm  0.001$&$  0.068\pm  0.002$&$  0.051\pm  0.001$&$  0.045\pm  0.001$&$  0.030\pm  0.000$&$  0.024\pm  0.001$&$  0.014\pm  0.001$\cr
[\ion{N}{2}]$\lambda6548$&$  0.208\pm  0.002$&$  0.109\pm  0.001$&$  0.063\pm  0.001$&$  0.034\pm  0.001$&$  0.022\pm  0.001$&$  0.014\pm  0.001$&$  0.010\pm  0.001$\cr
H$\alpha$&$  2.860\pm  0.015$&$  2.860\pm  0.014$&$  2.860\pm  0.008$&$  2.860\pm  0.006$&$  2.860\pm  0.006$&$  2.860\pm  0.010$&$  2.860\pm  0.014$\cr
[\ion{N}{2}]$\lambda6584$&$  0.626\pm  0.004$&$  0.354\pm  0.002$&$  0.201\pm  0.001$&$  0.118\pm  0.001$&$  0.066\pm  0.001$&$  0.045\pm  0.001$&$  0.025\pm  0.001$\cr
[\ion{S}{2}]$\lambda6717$&$  0.269\pm  0.002$&$  0.220\pm  0.002$&$  0.174\pm  0.001$&$  0.153\pm  0.001$&$  0.104\pm  0.001$&$  0.076\pm  0.001$&$  0.050\pm  0.001$\cr
[\ion{S}{2}]$\lambda6731$&$  0.223\pm  0.002$&$  0.172\pm  0.002$&$  0.132\pm  0.001$&$  0.116\pm  0.001$&$  0.080\pm  0.001$&$  0.057\pm  0.001$&$  0.037\pm  0.001$\cr
used&n&y&y&y&y&y&y\cr
$T_e$ Metallicity&  -&$  8.37\pm  0.02$&$  8.25\pm  0.01$&$  8.16\pm  0.009$&$  8.03\pm  0.006$&$  7.91\pm  0.006$&$  7.81\pm  0.008$\cr
\hline
\hline
\hline
\end{tabular}
\end{table*}

\subsection{Diagnostic Diagrams}

Figure~\ref{BPT}(a) shows the selection of the analogs of high-redshift galaxies and normal SDSS galaxies in the reference sample in the {\oiii/\hb} versus {\nii/\ha} BPT diagram. In addition, we compare the location of our local analogs and normal SDSS galaxies in other optical diagnostic diagrams in Figure~\ref{BPT}(b-e). We find that our analogs share the same regions with $z\sim2$ star-forming galaxies in all these diagnostic diagrams \citep{Steidel:2014aa,Shapley:2015aa,Sanders:2016aa}. Compared to normal SDSS galaxies, our analogs do not show offsets in the {\oiii/\hb} vs. {\siid/\ha} (S2-) BPT diagram (Figure~1(b)) and O32 vs. R23 diagram (Figure~1(d)), but do show shifts in the O32 vs. N2 and O32 vs. O3N2 diagrams (Figures~\ref{BPT}(e) and \ref{BPT}(f)).  We also find no significant offset between our analogs and local SDSS galaxies in the {\oiii/\hb} vs. {\oi/\ha} (O1-) BPT diagram (Figure \ref{BPT}(c)). The O1-BPT diagram has not been studied in $z\sim2$ galaxies, and our result can be tested in the future studies of $z\sim2$ galaxies.

\begin{figure}[]
\begin{center}
\includegraphics[width=1.0\columnwidth,angle=0]{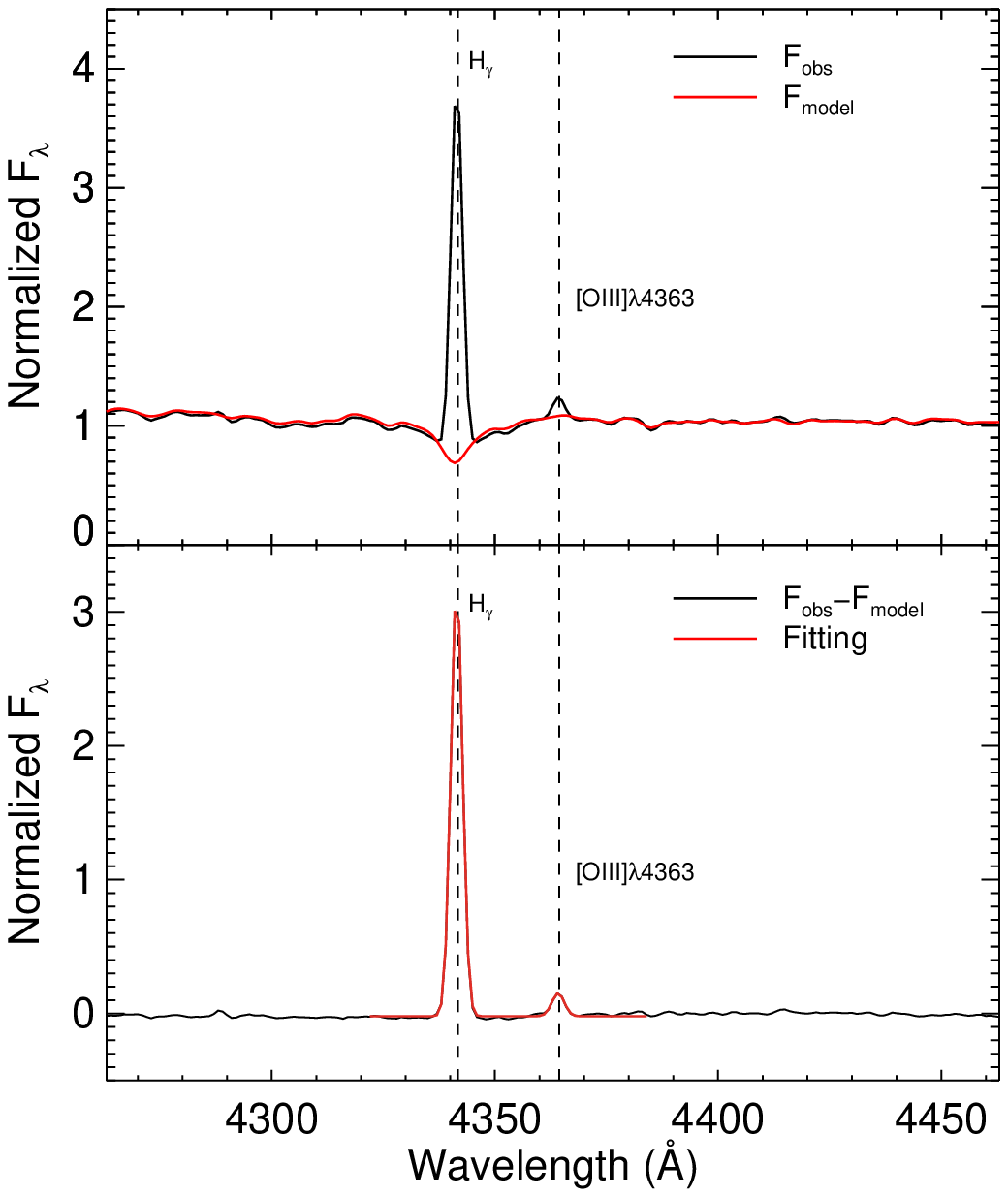}
\caption{Example of stacked spectra and the stellar continuum and emission line fitting in the wavelength range close to H$\gamma$ and {\oau}. {\it Top panel:} The black solid line represents an example of the stacked spectrum and the red line represents the best-fit stellar continuum for the stacked spectrum. {\it Bottom panel:} The black solid line represents the stellar-continuum subtracted spectrum and the red line represents the fitting result of H$\gamma$ and {\oau} emission line.
 \label{spec_example}}
\end{center}
\end{figure}

\section{Analysis}\label{analysis}
\subsection{Generating Stacked Spectra}
We cannot detect the weak {\oau} emission line in most of our individual spectra. To increase the signal-to-noise ratio, we combine the spectra and generate stacked spectra for both our local analogs and selected SDSS galaxies. We adopt the procedure that has been used in \citet{Andrews:2013aa} and \citet{Brown:2016aa}, except we carry out dust extinction correction before stacking the spectra. The procedure is as follows:
\begin{enumerate}
\item We obtain the reduced one-dimensional SDSS spectra from the SDSS DR9 data release. The spectra were taken using the multi-object, fiber-fed spectrographs \citep{Smee:2013aa} mounted on the 2.5m telescope of SDSS \citep{Gunn:2006aa}. The spectra have been reduced by the SDSS {\tt spectro2d} pipeline \citep{Stoughton:2002aa}.
\item Each of the SDSS galaxy spectra is corrected for the dust extinction using the Balmer decrement (\ha/\hb) assuming Case B recombination and the \citet{Cardelli:1989aa} dust extinction law.
\item Each SDSS spectrum is shifted to the rest frame based on the redshift from the MPA/JHU catalog.
\item Each of the SDSS spectra is resampled onto a wavelength grid in the $3700-7300${\AA} range with $\Delta\lambda=1${\AA} while maintaining flux conservation.
\item We normalize the spectra with the mean flux density in the $4400-4450${\AA} wavelength range.
\item We divide the local analogs and the selected SDSS galaxies separately into 0.25 dex bins in {\nii}/{\ha} from $\log$({\nii}/{\ha})$ =x$ to $x+0.25$, where $x=[-2.25, -2.00, -1.75, -1.50, -1.25, -1.00, -0.75]$. We use the N2$^{x+0.25}_{x}$ to represent the $\log$({\nii}/{\ha}) between $x$ and $x+0.25$ for the remainder of the paper.
\item We stack the spectra falling into each of the {\nii}/{\ha} bins using the mean flux density at each wavelength. 
\end{enumerate}
Following the above procedure, we generate a total of 14 stacked spectra: 7 for the local analogs of high-redshift galaxies and 7 for the SDSS reference sample.

\subsection{Subtracting the Stellar Component}
The underlying stellar absorption features in the stacked spectra affect the emission-line flux measurements. In particular, the H$\gamma$ stellar absorption feature is close to the {\oau} emission line. Therefore, to measure the line fluxes accurately, it is important to subtract the stellar continuum from the stacked spectra. We fit the stellar continuum of each stacked spectrum with a set of stellar synthesis models using the STARLIGHT stellar population synthesis code \citep{Cid-Fernandes:2005aa}. We mask out the emission-line region before fitting the spectra. The \citet{Bruzual:2003aa} stellar synthesis models with a Chabrier initial mass function \citep{Chabrier:2003wd} are adopted to fit the stacked spectra. These stellar synthesis models cover a broad range of metallicity ($0.05Z_\sun-2.5Z_\sun$) and stellar age ($1~\rm{Myr}-13$~Gyr). The upper panel of Figure~\ref{spec_example} shows an example of the observed stacked spectrum ($F_{\rm obs}$) and the model spectrum ($F_{\rm model}$) from the best-fit stellar synthesis model at the wavelength close to the H$\gamma$ and [\ion{O}{3}$]\lambda$4363 lines. The model spectrum follows the continuum of the observed spectrum and captures most of the spectral absorption features, suggesting our stellar synthesis model fitting process is robust. 

\subsection{Measuring Emission Line Flux}
We measure the following emission-line fluxes in the stellar continuum subtracted stacked spectra: {\oiid}, H$\gamma$, \oau, \hb, \ha, \niid, and {\siid}. We use the IDL MPFIT package to fit the emission lines by assuming that the emission-line profile follows the Gaussian distribution. A single Gaussian function is used to fit the emission lines that are well separated ($>30$~{\AA}) from other emission lines, including the {\hb}, [\ion{O}{3}]$\lambda$4959, {\oiii}. Two Gaussians are fit simultaneously to those pairs of lines closer than 30~{\AA}, including the H$\gamma$ and {\oau} pair, the {\oiid} doublet, and the {\siid} doublet. As for the {\ha} and {\niid} complex, three Gaussians are used to fit simultaneously. In the fitting process, the central wavelength is fixed to the vacuum wavelength of the emission lines. We adopt the standard deviation of  the continuum region close to ($<100$~\AA) the emission-line-fitting regions as the 1$\sigma$ noise of the spectra. We carry out Monte Carlo simulations to estimate the uncertainties of line fluxes. A thousand simulated spectra are generated. The flux densities of these simulated spectra at each wavelength follow a Gaussian distribution whose center is at the actual flux density, and the standard derivation is consistent with 1$\sigma$ error at that wavelength. The bottom panel of Figure~\ref{spec_example} shows the model-subtracted spectrum ($F_{\rm obs}-F_{\rm model}$) and best Gaussian fit for the H$\gamma$ and [\ion{O}{3}]$\lambda$4363 line, which demonstrates that our line-fitting is robust. Tables~\ref{tab:SDSS_flux} and \ref{tab:analog_flux} summarize the emission-line fluxes in each stacked spectrum. 

Figure~\ref{BPT}(a) shows the location of the stacked spectrum on the {\oiii/\hb} versus the {\nii/\ha} BPT diagram. We find that all the stacked spectra closely trace the local star-forming sequence and high-redshift star-forming sequence on the BPT diagram except the stacked spectrum of normal SDSS galaxies in the N2$^{-2.25}_{-2.00}$ bin. However, only two galaxies in the SDSS reference galaxy sample fall into the N2$^{-2.25}_{-2.00}$ bin, and both of them are located above the local star-forming sequence, biasing the stack spectrum toward the high-redshift star-forming sequence. Therefore, this data point is discarded for further analysis.

To achieve reliable direct oxygen abundance measurements, we only use the stacked spectra with S/Ns {\oau} greater than ten ({$\rm S/N>10$}) for further analysis.

\subsection{Measuring the Direct Oxygen Abundance}
We adopt the \citet{Izotov:2006ab} recipe to derive the direct $T_e$ oxygen abundance. This method builds up a relation between the {\oiiid}/{\oau} ratio and the electron temperature in the O$^{++}$ zone ($T_e$(\ion{O}{3})). This relation also weakly depends on the electron density, which can be estimated from the {\siid} doublet ratio. We do not have directly measured electron temperatures in the O$^+$ zone ($T_e$(\ion{O}{2})). We adopt a relation between $T_e$(\ion{O}{3}) and $T_e$(\ion{O}{2}) from \citet{Garnett:1992aa} and \citet{Campbell:1986aa} as follow: $T_e$(\ion{O}{2})=0.7$T_e$(\ion{O}{3})+3000K. We estimate the O$^{++}$ abundance using the $T_e$(\ion{O}{3}) temperature and {\oiiid/\hb} ratio and O$^+$ abundance using the $T_e$(\ion{O}{2}) temperature, {\oiid/\hb} and electron density. We compute the oxygen abundance by summing the O$^{++}$ abundance and the O$^{+}$  abundance. We refer the reader to section~3.1 in \citet{Izotov:2006ab} for more details.

\subsection{Interstellar Medium Conditions}
We study the ISM conditions, including the ionization parameter and electron density in the SDSS reference galaxies and local analogs of high-redshift galaxies. We estimate the 
ionization parameter in stacked spectrum in each N2 bin by adopting the recipe that first introduced in \citet{Kobulnicky:2004aa} \citep[see also][]{Kewley:2008aa}. This method is based on the \citet{Kewley:2002fk} photoionization models. The ionization parameter is calculated by primarily using the O32. For a fixed O32, the oxygen abundance has a secondary effect on the ionization parameter measurement. So we further correct for the ionization parameter based on the oxygen abundance from the R23 and N2O2. We refer readers to section A2.3 in \citet{Kewley:2008aa} for more details on the ionization parameter measurements. We estimate the electron density in each of the stacked spectra based on the [\ion{S}{2}]$\lambda$6717/[\ion{S}{2}]$\lambda$6731 ratio by adopting the {\tt nebular.temden} routine in IRAF\citep{Shaw:1995aa}. Tables~\ref{tab:SDSS_properties} and \ref{tab:analog_properties} summarize the results of the ionization parameters and electron densities in the SDSS reference galaxies and local analogs. We also list the median stellar mass\footnote{The stellar mass measurements are adopted from \citet{Bian:2016aa}} and SFR for each N2 bin in Tables~\ref{tab:SDSS_properties} and \ref{tab:analog_properties}. The stellar masses of the local analogs and SDSS reference galaxies are consistent with each other in each of the N2 bins. The local analogs have higher ionization parameters ($\sim0.3$~dex), higher SFRs, and higher electron densities (a factor of $\sim5$) than SDSS reference galaxies in all N2 bins except the  N$_{-2.00}^{-2.25}$ bin, \footnote{In the  N2$_{-2.00}^{-2.25}$ bin, the selected SDSS reference galaxies fall onto the high-redshift BPT star-forming locus.} suggesting that the ISM conditions change significantly between the SDSS reference galaxies and local analogs. 

\begin{figure*}[]
\begin{center}
\includegraphics[width=1.8\columnwidth]{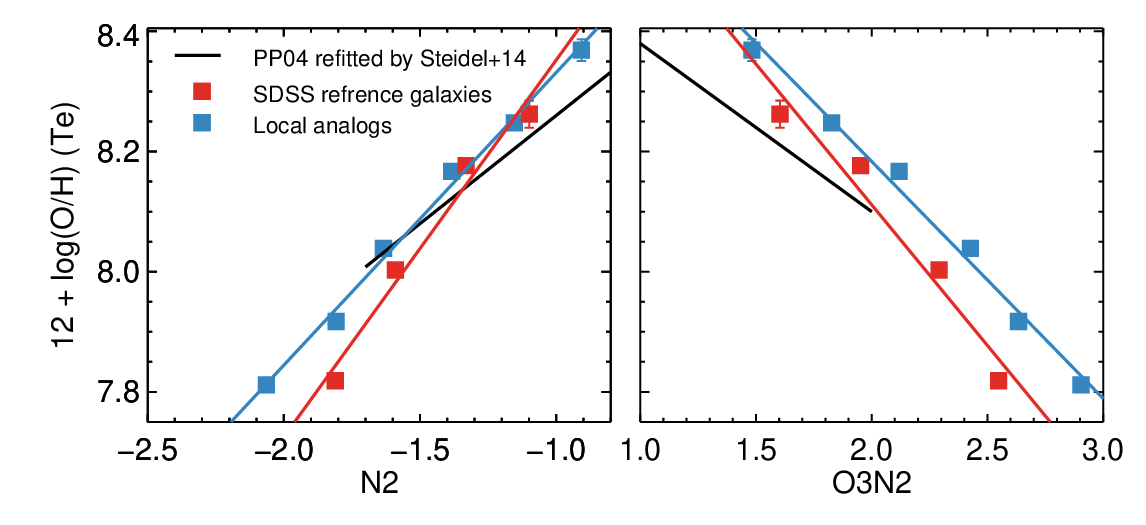}
\caption{Relations between the direct oxygen abundance and N2 (left) and O3N2 (right) indices in the local analogs of high-redshift galaxies and the SDSS reference galaxies. The blue and red filled squares represent the local analogs and the SDSS reference galaxies, respectively. The blue and red lines represent the best fit of the relations between the direct oxygen abundance and N2 and O3N2 indices for the local analogs and the SDSS reference galaxies, respectively. The black line represents the empirical relation between the oxygen abundance and N2 and O3N2 indices from \citet{Steidel:2014aa}, who refitted the N2 and O3N2 metallicity calibrations of PP04 in the range of $-1.7<\rm{N2}<-0.3$ and $-0.4<\rm{O3N2}<2.1$
\label{N2_O3N2}}
\end{center}
\end{figure*}

\begin{table*}[h!]
\centering
\caption{The Properties of the SDSS Reference Galaxies.\label{tab:SDSS_properties}}
\begin{tabular}{ccccccccccccccccccc}
 \hline
Bin&\#\footnote{Number of galaxies for a given N2 bin} &SFR\footnote{The uncertainty in the SFR represents
the 16th and 84th percentiles of the SFR distribution in a given N2 bin} &Stellar Mass\footnote{The uncertainty in the stellar mass represents
the 16th and 84th percentiles of the stellar mass distribution in a given N2 bin}  & $\log q$\footnote{Ionization parameter derived from the stacked spectrum}& $n_e$ \footnote{Electron density derived from the stacked spectrum}&$T_e$ Metallicity\\
\hline
&&$M_\sun$~yr$^{-1}$&$\gmass$&cm~s$^{-1}$&cm$^{-3}$&$12+\log(\rm{O/H})$\\
\hline
\hline
      N2$_{-0.75}^{-0.50}$ &15927 &        $  1.1\substack{+  1.7\\-  0.7}$&        $  9.6\substack{+  0.3\\-  0.3}$ &                    $ 7.39\pm 0.02$ &$     5\substack{+     9\\-     2}$&-\\
      N2$_{-1.00}^{-0.75}$ & 4491 &        $  0.9\substack{+  2.2\\-  0.6}$&        $  9.2\substack{+  0.3\\-  0.3}$ &                    $ 7.57\pm 0.03$ &$    16\substack{+    18\\-    14}$&-\\
      N2$_{-1.25}^{-1.00}$ & 1595 &        $  0.7\substack{+  1.9\\-  0.5}$&        $  8.9\substack{+  0.4\\-  0.5}$ &                    $ 7.72\pm 0.04$ &$    22\substack{+    25\\-    19}$&$8.26\pm0.02$\\
      N2$_{-1.50}^{-1.25}$ &  333 &        $  0.5\substack{+  1.2\\-  0.3}$&        $  8.3\substack{+  0.5\\-  0.5}$ &                    $ 7.74\pm 0.03$ &$    21\substack{+    25\\-    17}$&$8.18\pm0.009$\\
      N2$_{-1.75}^{-1.50}$ &   63 &        $  0.5\substack{+  1.6\\-  0.3}$&        $  7.8\substack{+  0.5\\-  0.4}$ &                    $ 7.92\pm 0.07$ &$    34\substack{+    42\\-    25}$&$8.00\pm0.008$\\
      N2$_{-2.00}^{-1.75}$ &   17 &        $  0.5\substack{+  0.9\\-  0.2}$&        $  7.7\substack{+  0.2\\-  0.5}$ &                    $ 8.10\pm 0.11$ &$    51\substack{+    72\\-    31}$&$7.80\pm0.006$\\
      N2$_{-2.25}^{-2.00}$ &    2 &        $  0.6\substack{+  1.5\\-  0.2}$&        $  7.3\substack{+  0.4\\-  0.4}$ &                    $ 8.44\pm 0.15$ &$   167\substack{+   204\\-   132}$&-\\
                   \hline
 \end{tabular}


\centering
\caption{The properties of the local analogs of high-redshift galaxies.\label{tab:analog_properties}}
\begin{tabular}{ccccccccccccccccccc}
 \hline
Bin&\#\footnote{Number of galaxies for a given N2 bin} &SFR\footnote{The uncertainty in the SFR represents
the 16th and 84th percentiles of the SFR distribution in a given N2 bin} &Stellar Mass \footnote{The uncertainty in the stellar mass represents
the 16th and 84th percentiles of the stellar mass distribution in a given N2 bin}   & $\log q$\footnote{Ionization parameter derived from the stacked spectrum}& $n_e$ \footnote{Electron density derived from the stacked spectrum}\\
\hline
&&$M_\sun$~yr$^{-1}$&$\gmass$&cm~s$^{-1}$&cm$^{-3}$&$12+\log(\rm{O/H})$\\
\hline
\hline
      N2$_{-0.75}^{-0.50}$ &   45 &        $  3.4\substack{+ 12.3\\-  2.9}$&        $  9.6\substack{+  0.3\\-  0.7}$ &                    $ 7.77\pm 0.09$  &$   255\substack{+   277\\-   235}$&-\\
      N2$_{-1.00}^{-0.75}$ &   69 &        $ 13.5\substack{+ 20.3\\- 11.2}$&        $  9.2\substack{+  0.3\\-  0.3}$ &                    $ 7.88\pm 0.18$  &$   159\substack{+   177\\-   141}$&$8.37\pm0.02$\\
      N2$_{-1.25}^{-1.00}$ &  110 &        $  8.0\substack{+ 20.3\\-  6.5}$&        $  8.9\substack{+  0.4\\-  0.6}$ &                    $ 7.97\pm 0.16$  &$   110\substack{+   119\\-   102}$&$8.25\pm0.01$\\
      N2$_{-1.50}^{-1.25}$ &   99 &        $  2.6\substack{+  6.6\\-  2.0}$&        $  8.4\substack{+  0.4\\-  0.5}$ &                    $ 7.95\pm 0.08$  &$   111\substack{+   124\\-    98}$&$8.16\pm0.009$\\
      N2$_{-1.75}^{-1.50}$ &   44 &        $  1.2\substack{+  1.7\\-  0.8}$&        $  8.0\substack{+  0.3\\-  0.3}$ &                    $ 8.14\pm 0.10$  &$   132\substack{+   156\\-   109}$&$8.03\pm0.006$\\
      N2$_{-2.00}^{-1.75}$ &   11 &        $  1.8\substack{+  2.1\\-  1.7}$&        $  7.8\substack{+  0.2\\-  0.4}$ &                    $ 8.33\pm 0.14$  &$    95\substack{+   137\\-    56}$&$7.91\pm0.006$\\
      N2$_{-2.25}^{-2.00}$ &    2 &        $  0.4\substack{+  1.7\\-  0.2}$&        $  7.3\substack{+  0.4\\-  0.1}$ &                    $ 8.50\pm 0.14$  &$    97\substack{+   144\\-    54}$&$7.81\pm0.008$\\
            \hline
 \end{tabular}

\end{table*}

\section{Results}\label{results}

\subsection{N2 and O3N2 Empirical Metallicity Calibrations}
Here, we investigate the N2 and O3N2 empirical metallicity calibrations in the local analogs and the SDSS reference galaxies by using the direct oxygen abundance and line ratios measured in the stacked spectrum. Figure~\ref{N2_O3N2} shows the relation between the direct $T_e$ oxygen abundance and the N2 and O3N2 indices in the local analogs of high-redshift galaxies (blue squares) and the local normal star-forming galaxies (red squares). The relation between the direct oxygen abundance and the N2 and O3N2 indices is fit with a linear equation for both the local analogs of high-redshift galaxies and SDSS reference galaxies. The results for the least-squares fit for our local analogs as follows (blue solid lines in Figure~\ref{N2_O3N2}):
\begin{equation}
12+\log(\rm{{O}/{H}}) = 8.82+0.49\times\rm{N2},
\label{highz_N2}
\end{equation}
\begin{equation}
12+\log(\rm{{O}/{H}})=8.97-0.39\times \rm{O3N2},
\label{highz_O3N2}
\end{equation}
The results for the SDSS reference galaxies are as follows (red solid lines in Figure~\ref{N2_O3N2}):
\begin{equation}
12+\log(\rm{{O}/{H}}) =  8.98+0.63\times\rm{N2},
\end{equation}
\begin{equation}
12+\log(\rm{{O}/{H}}) = 9.05 -0.47\times \rm{O3N2},
\end{equation}
where $7.8<12+\log(\rm{{O}/{H}})<8.4$

In Figure~\ref{N2_O3N2}, we compare our N2 and O3N2 calibrations in the local analogs, the SDSS reference galaxies, and the local \ion{H}{2} regions. For the local \ion{H}{2} region, we use the metallicity calibrations from \citet{Steidel:2014aa} rather than those from PP04. \citet{Steidel:2014aa} repeated the fits to N2 and O3N2 metallicity calibrations in PP04, but only focused on the \ion{H}{2} regions with direct $T_e$ metallicity and in the low-metallicity range ($-1.7<\rm{N2}<-0.3$ and $-0.4<\rm{O3N2}<2.1$). Thus, the metallicity calibrations from \citet{Steidel:2014aa} provide better comparisons with the relations in this work. The new N2 and O3N2 calibrations based on the local analogs are not consistent with those found in both the SDSS reference galaxies and the local \ion{H}{2} region PP04. The discrepancies between the SDSS reference galaxies and the local \ion{H}{2} regions are mainly caused by the contribution of diffused ionized gas emission in the SDSS reference galaxies \citep{Sanders:2017aa} and slightly different recipes for deriving the direct $T_e$ metallicity. The contribution of the diffused ionized gas emission in the local analogs is negligible due to their high H$\alpha$ surface density \citep{Zhang:2017aa}. The direct $T_e$-based metallicity in the local analogs is systematically higher than that in the SDSS reference galaxies for a given N2 or O3N2 in the metallicity range of $7.8<12+\log(\rm O/\rm H)<8.4$. This discrepancy is mainly due to the changes of ISM conditions between our analogs and normal SDSS star-forming galaxies. The high ionization parameter and electron density increase the {\nii} and {\oiii} line fluxes for a given metallicity, which causes the lower N2 and higher O3N2 values in the analogs \citep[][see Section~\ref{sec:model} for a further discussion]{Dopita:2016aa}.

\begin{figure*}[]
\begin{center}
\includegraphics[width=1.7\columnwidth]{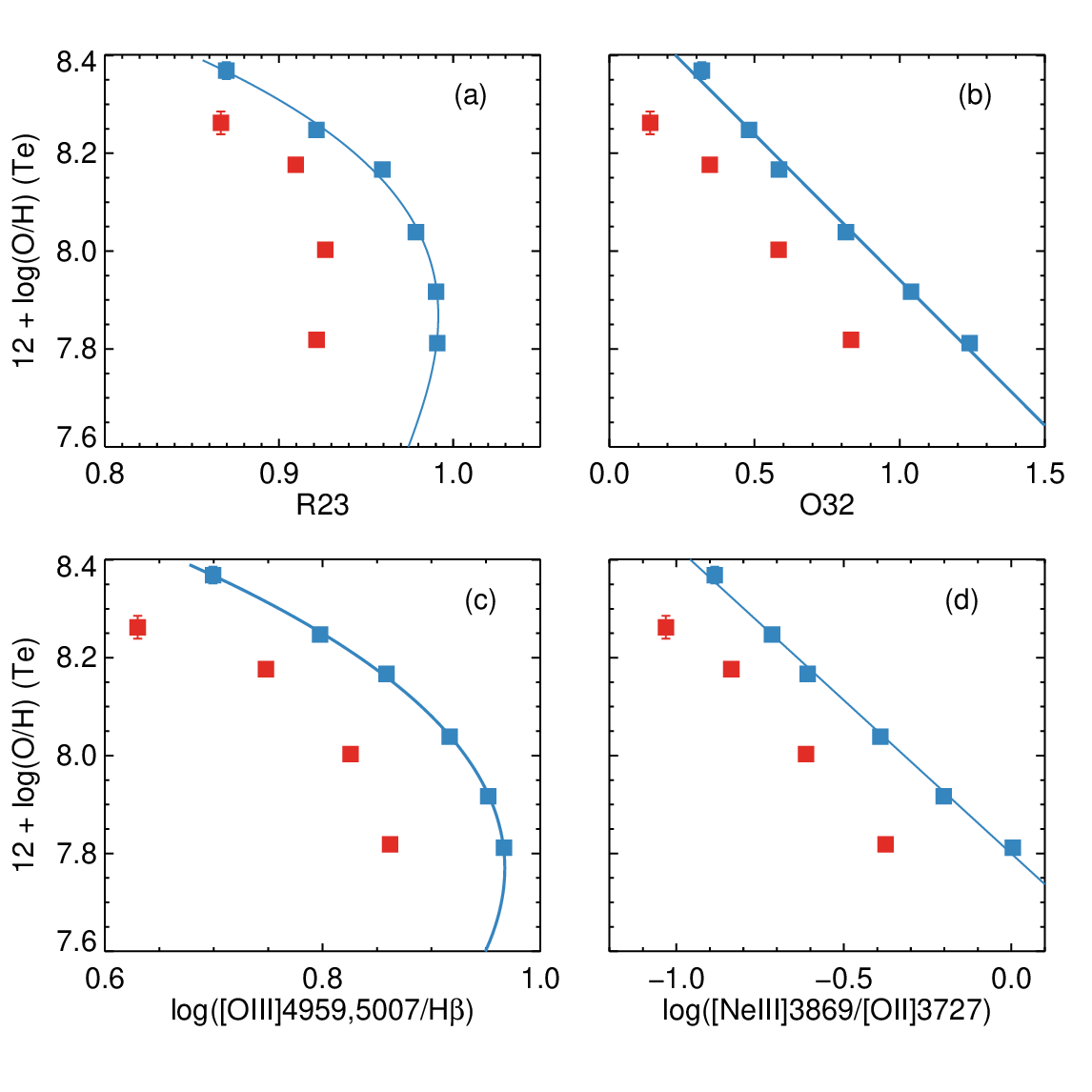}
\caption{Relation between the direct oxygen abundance and R23, O32,  $\log$\oiiid/\hb, and $\log$({\neiii}/{\oii}) indices. The blue filled squares represent the local analogs of high-redshift galaxies, and the red filled squares represent the SDSS reference galaxies.\label{R23_O32} }
\end{center}
\end{figure*}

\subsection{Empirical Metallicity Calibrations for R23, O32, {\oiiid/\hb}, and {\rm[\ion{Ne}{3}]/[\ion{O}{2}]}}
The {\ha} and {\nii} emission lines move out of the $K$-band atmospheric transmission window for galaxies at $z>2.65$. Therefore, metallicity diagnostics based on the emission lines at shorter wavelengths are essential for galaxies at $z>2.65$. We study the empirical metallicity calibrations for the R23, O32 and {\oiiid/\hb}, and [\ion{Ne}{3}]$\lambda$3869/[\ion{O}{2}] ratios. The [\ion{O}{2}], [\ion{O}{3}], {\hb}, [\ion{Ne}{3}] lines can be accessed in the $H$- and $K$-bands for galaxies at $z\sim3.5$, and the [\ion{Ne}{3}]/[\ion{O}{2}] ratio can be measured in galaxies with redshift as high as $z=5.0$.  These lines provide a powerful tool to measure the metallicity in galaxies at high redshift by assuming the ISM conditions did not change dramatically from $z=2$ to $z=5$\footnote{Given the evidence that the sSFR plays an important role to regulate the ISM conditions in galaxies \citep[e.g.,][]{Bian:2016aa} and the sSFR does not change dramatically from $z=2$ to $z=5$ \citep[e.g.][]{Stark:2013aa}}.

Figure~\ref{R23_O32} shows the relation between the $T_e$ metallicity and the four metallicity diagnostics measured previously. We find that the our local analogs and normal SDSS galaxies do not follow the same relation in all four relations. For a given metallicity, the R23, O32, {\oiiid/\hb}, and [\ion{Ne}{3}]/[\ion{O}{2}] ratios in the local analogs are higher that those in the normal SDSS galaxies. We fit the relations between the metallicity and R23, O32, {\oiiid/\hb}, and [\ion{Ne}{3}]/[\ion{O}{2}] ratios for the local analogs of high-redshift galaxies. These relations are suitable for the metallicity range of $7.8<12+\log(\rm O/\rm H)<8.4$.


For the R23-Z relation, our data only cover the R23 upper branch and the transition zone of the R23 upper and lower branches. We fit the upper branch of the R23-Z relation using a third-order polynomial:
\begin{equation}
y= 138.0430-54.8284x+7.2954x^2-0.32293x^3
\end{equation}
where $y=R23$ and $x=12+\log(\rm{O/H})$. 
It is worth noting that the metallicity is not very sensitive to the R23 values in the metallicity range of $7.8<12+\log(\rm O/\rm H)<8.4$, because most of the data points are located in the transition zone of the R23 upper and lower branches.

For the O32-Z relation, we fit the data points using a linear equation:
\begin{equation}
12+\log(\rm{{O}/{H}}) = 8.54-0.59\times\rm{O32}.
\end{equation}


For the {\oiiid/\hb}-Z relation, we fit the data points using a third-order polynomial:
\begin{equation}
y=43.9836-21.6211x+3.4277x^2-0.1747x^3
\end{equation}
where $y=${\oiiid/\hb} and $x=12+\log(\rm{O/H})$. 

For the [NeIII]/[OII]-Z relation, we fit the data points using a linear equation:
\begin{equation}
12+\log(\rm{{O}/{H}}) =  7.80-0.63\times\rm{\log([NeIII]/[OII])}.
\end{equation}


\begin{figure}[h]
\begin{center}
\includegraphics[width=1.0\columnwidth]{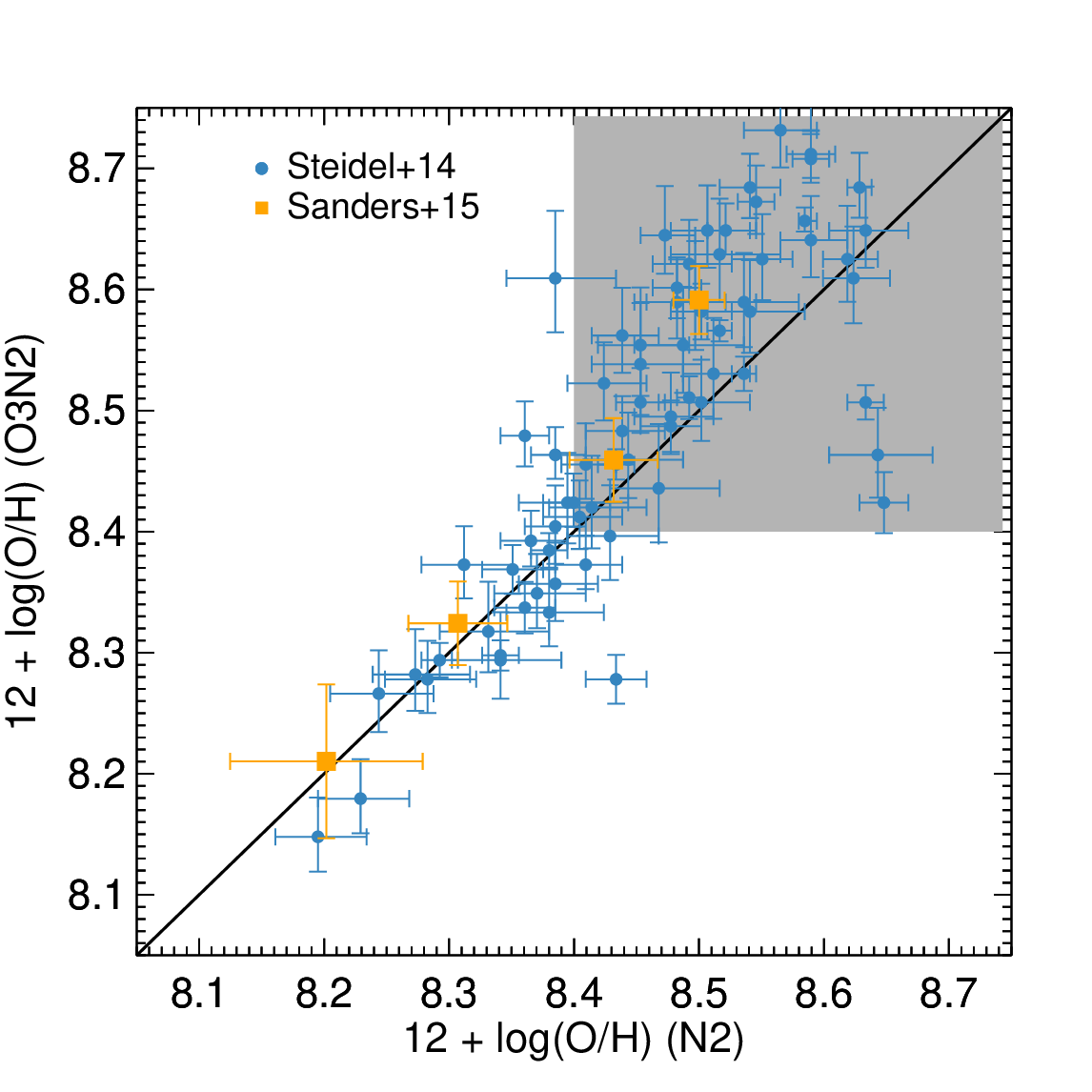}
\caption{Comparison between the N2-based metallicity and O3N2-based metallicity in $z\sim2$ star-forming galaxies using the BKD18 (this work) calibrations in this study. 
The blue filled circles represent the individual spectra of $z\sim2$ UV-selected star-forming galaxies from \citet[][S14]{Steidel:2014aa}. The orange filled circles represent the stacked spectra of $z\sim2$ mass-selected star-forming galaxies in four mass bins from \citet[][S15]{Sanders:2015aa}. The shaded region denotes the metallicity range in which the BKD18 calibrations are not valid. The BKD18 calibrations are valid in the metallicity range $7.8<12+\log(\rm O/\rm H)<8.4$. In this metallicity range, the N2- and O3N2-based metallicity are consistent with each other in both UV-selected and mass-selected star-forming galaxies.
\label{N2_O3N2_compare}}
\end{center}
\end{figure}

\begin{figure*}
\begin{center}
\includegraphics[width=2\columnwidth]{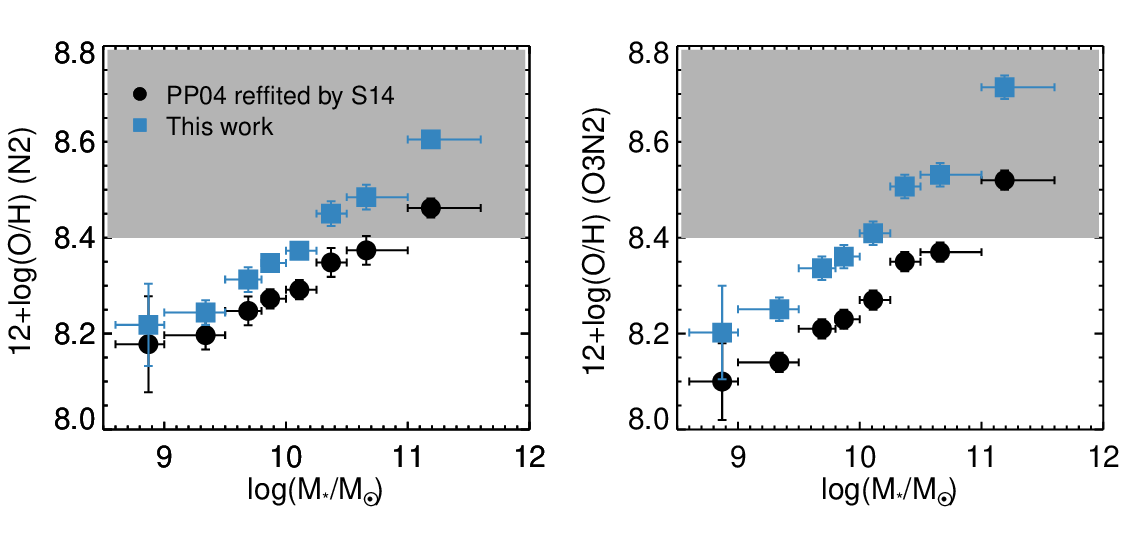}
\caption{Mass-metallicity relation in $z\sim2$ UV-selected star-forming galaxies \citep{Steidel:2016aa} based on the N2 (left) and O3N2 (right) diagnostics using the PP04 relation that was refitted by \citet{Steidel:2014aa} (black data points) and the BKD18 (blue data points) calibrations. The shaded region represents the metallicity ranges that are out of the BKD18 calibration fitting range. \label{mzr}}
\end{center}
\end{figure*}

\begin{figure}[h]
\begin{center}
\includegraphics[width=1.0\columnwidth]{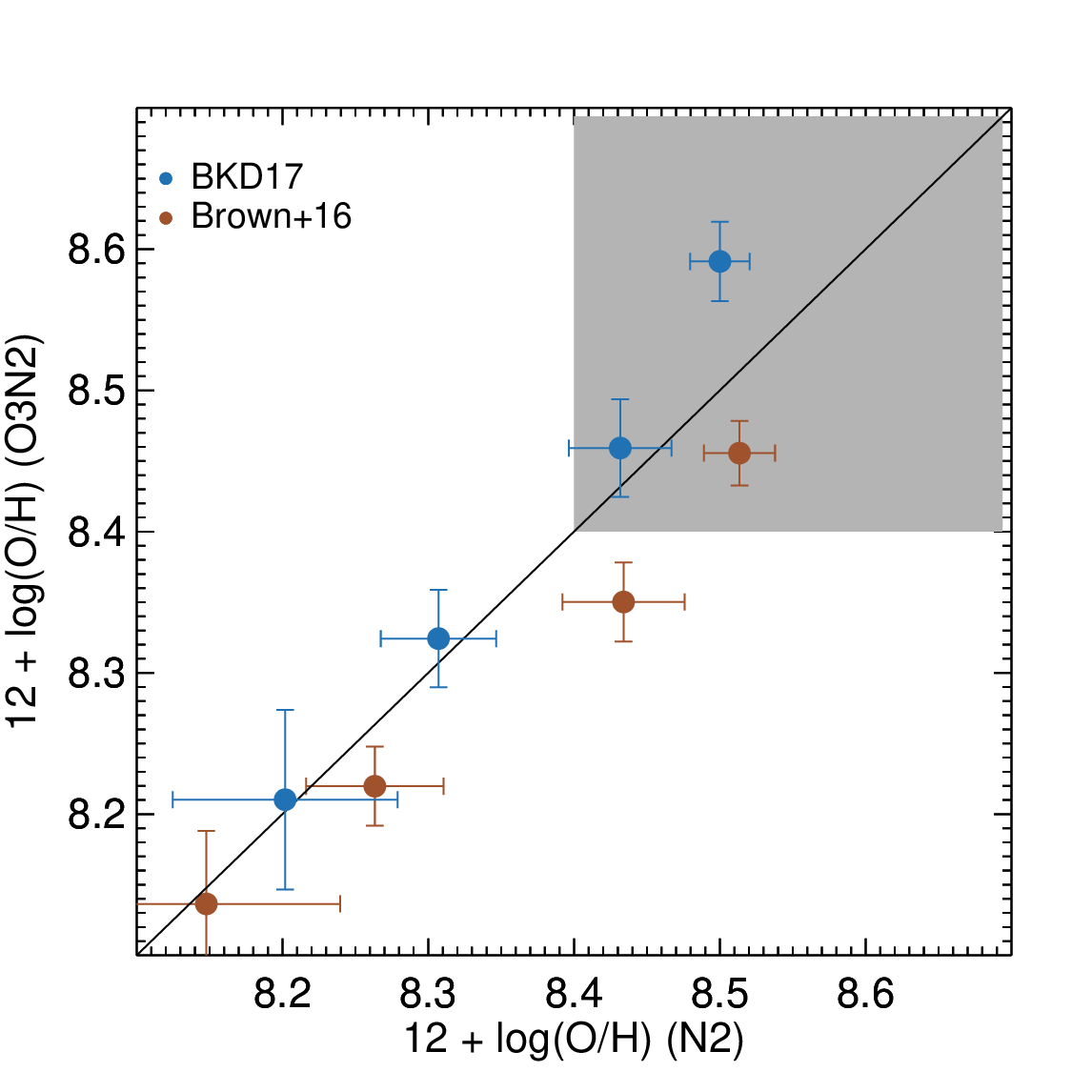}
\caption{Comparison between the N2-based metallicity and O3N2-based metallicity in $z\sim2$ star-forming galaxies using the BKD18 (blue data points) and \citet[][brown data points]{Brown:2016aa} calibrations. The solid line represents the one-to-one relation between the N2 and O3N2 metallicity. The shaded region represents the metallicity range in which the BKD18 calibrations are not valid. The BKD18 calibrations are valid in the metallicity range $7.8<12+\log(\rm O/\rm H)<8.4$. In this metallicity range, the N2 and O3N2 metallicities in the BKD18 calibrations are more consistent with each other than the \citet{Brown:2016aa} calibration.\label{cal_comparison}}
\end{center}
\end{figure}

\section{Discussion}\label{discussion}
In this section, we refer to our new metallicity calibration based on the local analogs of high-redshift galaxies as the BKD18 calibration to simplify the expression. 

\subsection{Implications for High-redshift Galaxy Metallicity Measurements}
The PP04 metallicity diagnostics cause a discrepancy between the N2 and O3N2 metallicities in $z\sim2$ star-forming galaxies. The O3N2-based metallicity is systematically smaller than the N2-based metallicity by $0.10-0.15$~dex \citep[e.g.,][]{Hainline:2009aa,Bian:2010vn,Newman:2014aa,Steidel:2014aa,Sanders:2015aa}. This discrepancy is primarily due to the offset between the local and $z\sim2$ star-forming galaxies in the {\oiii/\hb} and {\nii/\ha} BPT diagram. Various physical mechanisms have been proposed to interpret this offset, including changes of ionization parameters, electron densities, radiation fields, and nitrogen-to-oxygen ratio over cosmic time. These proposed physical mechanisms are crucial input parameters for photoionization models and play an important role for the metallicity estimations. Our stacked spectra also shed light on what is the major physical mechanism(s) to drive the offset between high-redshift and low-redshift galaxies on the BPT diagram (F. Bian in preperation). 
 
To test our new calibrations in high-redshift galaxies, we apply the BKD18 N2 and O3N2 empirical metallicity calibrations (equations \ref{highz_N2} and \ref{highz_O3N2}) to two samples of $z\sim2$ star-forming galaxies. One sample is selected based on their UV emission properties, UV-selected star-forming galaxies from \citet{Steidel:2014aa}, and the other is selected based on their stellar mass, mass-selected star-forming galaxies from \citet{Sanders:2015aa}. The local analog selection criteria in this work are based on the location of the UV-selected galaxies on the BPT diagram. Therefore, the ISM conditions in our local analogs are more representative to those in the UV-selected galaxies, and the BKD18 metallicity calibration is more suitable for this type of galaxy. Figure~\ref{N2_O3N2_compare} shows the comparison between the N2- and O3N2-based metallicities in these UV-selected galaxies (blue filled circles) using our new metallicity calibrations. In the metallicity range of $8.1<\log(\rm{O/H})<8.4$ (shaded region in Figure~\ref{N2_O3N2_compare}), the mean difference between the BKD18 N2- and O3N2-metallicity is -0.03 dex. For a comparison, the mean difference between the PP04 N2- and O3N2-metallicity is 0.13 dex. The BKD18 calibrations successfully solve the discrepancy between N2- and O3N2-based metallicity when applying local PP04 calibrations to high-redshift galaxies \citep[e.g.,][]{Steidel:2014aa}. The UV-selected star-forming galaxies may only represent $50\%-70\%$ of the whole population of high-redshift star-forming galaxies \citep{Reddy:2005aa}. Therefore, we also apply our metallicity calibrations to a sample of mass-selected galaxies \citep{Sanders:2015aa}, which is a more representative sample of $z\sim2$ star-forming galaxies. \citet{Shapley:2015aa} found that the [\ion{O}{3}]/H$\beta$ ratio in mass-selected star-forming galaxies is not as high as UV-selected star-forming galaxies at $z\sim2$, i.e., the BPT locus of mass-selected galaxies is lower than the BPT locus of UV-selected galaxies at $z\sim2$. Our local analogs, therefore, may not fully resemble the ISM conditions of mass-selected galaxies, which could introduce systematical uncertainty when applying the new calibrations to mass-selected galaxies. In Figure~\ref{N2_O3N2_compare}, we plot the new N2- and O3N2-based metallicity based on the four stacked spectra of the mass-selected galaxies from \citet{Sanders:2016aa}.  In the metallicity range of $8.1<\log(\rm{O/H})<8.4$, the mean difference between the BKD18 N2- and O3N2-metallicity is -0.02 dex for mass-selected star-forming galaxies. For comparison, 
the mean difference between the PP04 N2- and O3N2-metallicity is 0.08 dex. Though the $z\sim2$ mass-selected galaxies is slight offset from the $z\sim2$ UV-selected galaxies on the BPT diagram \citep[e.g., Figure 1(a) in][]{Bian:2017aa}, the BKD18 calibrations are also suitable for $z\sim2$ mass-selected galaxies. 

It is worth noting that the N2 metallicity becomes systematically larger than the O3N2 metallicity in the BKD18 metallicity calibration, when $12+\log(\rm{O/H})>8.4$. The discrepancy increases when metallicity becomes larger. At $12+\log(\rm{O/H})=8.6$, the O3N2 metallicity is about 0.1 dex higher than N2 metallicity. This discrepancy suggests that the BKD18 calibration is not suitable for high-metallicity ($12+\log(\rm{O/H}>8.4$) cases. We do not recommend using the BKD18 calibration out of the metallicity range of $7.8<12+\log(\rm{O/H})<8.4$. It is difficult to use our local analogs to study the empirical calibration at $12+\log(\rm{O/H}) >8.4$. The direct $T_e$ method becomes insensitive (or saturated) to the high-metallicity ($12+\log({\rm O/H}) > 8.5$) galaxies, because {\oau} is only emitted in the hottest nebulae, which biases toward the low-metallicity region. Therefore, the direct $T_e$ method underestimates the oxygen abundance in such galaxies. 

The BKD18 metallicity calibrations provide a useful tool to study the mass-metallicity relation in high-redshift galaxies, particularly at the low-metallicity end. We measure metallicities in a sample of $z\sim2$ star-forming galaxies from \citet{Steidel:2014aa} using N2 and O3N2 indicators. Figure~\ref{mzr} shows the comparison between the mass-metallicity relation based on the BKD18 calibrations (blue squares) and that based on the PP04 calibrations (black circles). In the metallicity range of $8.0<Z<8.4$, we find that in both the N2 and  O3N2 based- metallicities that the BKD18 calibrations are systematically higher than those based on the PP04 calibrations by 0.05 to 0.1 dex.

 \begin{figure*}
\begin{center}
\includegraphics[width=2.2\columnwidth]{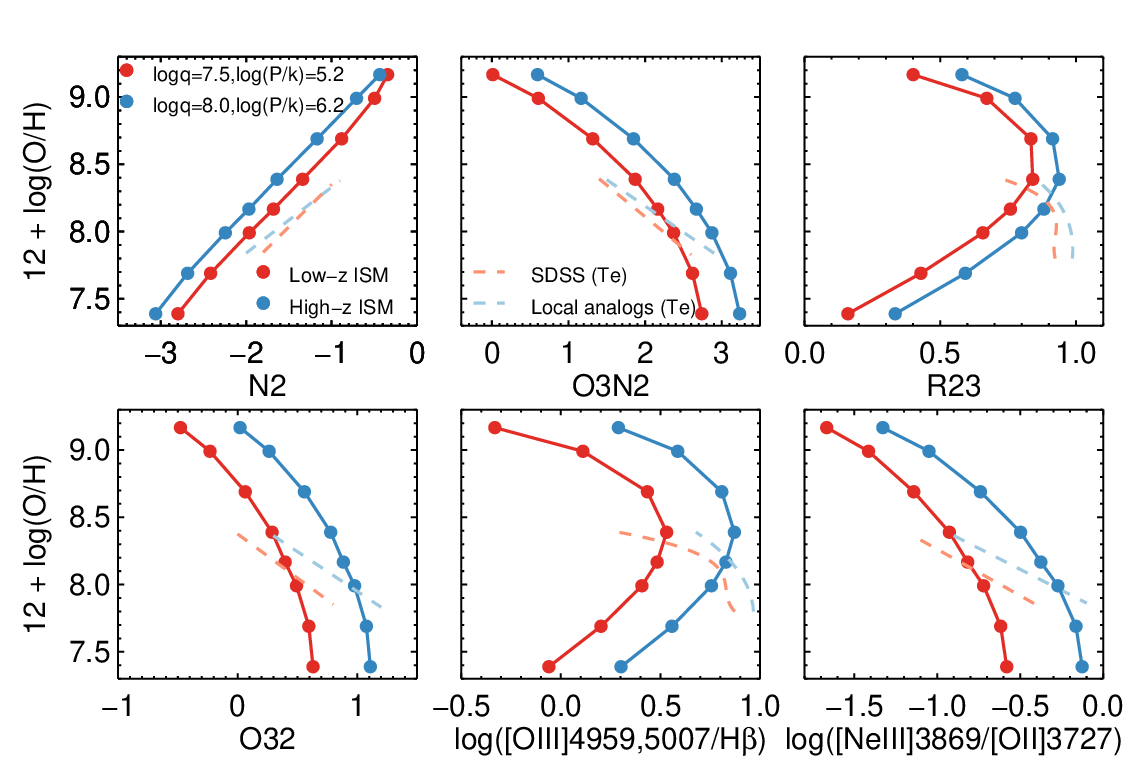}
\caption{Relations between the strong emission-line ratios and oxygen abundance from MAPPINGs photoionization models. Two sets of model grids are displayed. The red filled circles denote the photoionization models with ISM pressure $\log(P/k)=5.2$ (electron density of $n_e\sim10$) and photoionization parameter of $\log q=7.5$, which are representative values found in local star-forming galaxies. The blue filled circles denote the photoionization models with ISM  pressure $\log(P/k)=6.2$ (electron density of $n_e\sim100$) and photoionization parameter of $\log q=8.0$, which are comparable to those in $z\sim2$ star-forming galaxies and our local analogs.The dashed lines represent the metallicity calibrations derived from the direct $T_e$ method in the SDSS reference galaxies (light red dashed line) and the local analogs of high-redshift galaxies (light blue dashed line).\label{model}}
\end{center}
\end{figure*}

\subsection{Comparison with other high-redshift empirical metallicity calibrations}
Recently, a few other studies have recalibrated the N2 and O3N2 metallicity for high-redshift galaxies \citep[e.g.,][]{Brown:2016aa,Cowie:2016aa}.  Most of these studies used galaxies that are selected by matching their global properties to high-redshift galaxies, such as sSFRs \citep{Brown:2016aa} or {\hb} luminosities \citep{Cowie:2016aa}. However, studies have shown that low-redshift galaxies selected by matching the global properties of high-redshift galaxies do not have the same ISM conditions as their high-redshift counterparts \citep[e.g.,][]{Shirazi:2014ab,Bian:2016aa}. Therefore, simply matching global properties may miss crucial information on the change of the ISM conditions, which play an important role in regulating the relation between metallicity and emission-line ratios. In Figure~\ref{cal_comparison}, we compare the N2- and O3N2-based metallicities in a sample of mass-selected star-forming galaxies at $z\sim2$ from \citet{Sanders:2015aa} using the BKD18 calibrations and the \citet{Brown:2016aa} calibration. We adopt the measurements of the sSFR in the table~1 of \citet{Sanders:2015aa}. The mean difference between the \citet{Brown:2016aa} calibration N2- and O3N2-metallicities is 0.05~dex. We find the BKD18 calibration results in the better consistency between the N2- and O3N2-metallicities compared with the \citet{Brown:2016aa} calibration in the metallicity range of $8.1-8.4$.


\subsection{Comparison with Photoionization Models}\label{sec:model}
We study the evolution of the metallicity calibrations across the cosmic time by using the MAPPINGs photoionization models \citep{Sutherland:1993aa,Dopita:2013aa}. We use the MAPPINGS IV code to generate the relation between the diagnostic line ratios and the oxygen abundance for two cases.  In the first case, the input ionization parameter is $\log q=7.5$ and the ISM pressure is $\log(P/k)=5.2$, which corresponds to an electron density of $n_e\approx10$~cm$^{-3}$. These parameters are comparable to those in the normal SDSS galaxies \citep[e.g.,][]{Dopita:2006aa,Nakajima:2014aa,Bian:2016aa}. In the second case, the input ionization parameter is $\log q=8.0$ and the ISM pressure is $\log(P/k)=6.2$, which corresponds to an electron density of $n_e\approx100$~cm$^{-3}$. These parameters are consistent with those in $z\sim2$ galaxies and our local analogs \citep[e.g.,][]{Nakajima:2014aa,Bian:2016aa,Kaasinen:2017aa}. Figure~\ref{model} shows the relations between the different diagnostic line ratios and the oxygen abundance derived from the photoionization models.  The metallicity calibrations between the high-redshift and low-redshift conditions derived by the photoionization models (solid lines in Figure~\ref{model}) follow the same trend as those between the local analogs and the SDSS reference galaxies (dashed lines in Figure~\ref{model}). This suggests that the different metallicity calibration relations between the high- and low-redshift galaxies are due to the changes of the ISM conditions across cosmic time.

Although the $T_e$ metallicity and theoretical metallicity show the same relative trend for the dependence of ISM conditions for a given strong line ratio, the absolute metallicity derived from the direct $T_e$ method is significantly smaller than that from photoionization models by about 0.5~dex \citep[see also][]{Kewley:2008aa,Kewley:2018aa}. This discrepancy is mainly due to the (1) the Inhomogeneous temperature distribution in \ion{H}{2} regions. The temperature fluctuation and/or gradients in \ion{H}{2} regions bias the electron temperature measurements to the high temperature (low-metallicity) regions, thus the direct $T_e$ method underestimates the oxygen abundance.  Futhuremore, (2) $\kappa$-distributions of electron energy in \ion{H}{2} regions are also a factor. \citet{Nicholls:2012aa} suggested that the electron energy in \ion{H}{2} regions follows the $\kappa$-distributions rather than the Maxwell-Boltzmann distribution, and the electron temperatures derived based on the Maxwell-Boltzmann distribution are significantly higher than those from $\kappa$-distributions \citep{Nicholls:2013aa}. Finally, (3) the relation between $t3$ (electron temperature in O$^{++}$ zone) and $t2$ (electron temperature in O$^+$ zone) also plays a role. In this study, the $T_e$(\ion{O}{2})-$T_e$(\ion{O}{3}) relation from \citet{Campbell:1986aa} is adopted for a better comparison with previous studies. However, it is different from the $T_e$(\ion{O}{2})-$T_e$(\ion{O}{3}) relation in the MAPPINGS photoionization models \citep[$T_e$(\ion{O}{2})=0.685$T_e$(\ion{O}{3})+2100K][]{Dopita:2013aa}). The metallicities derived based on  the $T_e$(\ion{O}{2})-$T_e$(\ion{O}{3}) in the MAPPINGS photoionization models are 0.1-0.2 dex higher than the metallicities in this study. Meanwhile, the uncertainties of the input parameters in the photoionization models, particularly the shape of ionizing radiation field \citep[e.g.,][]{Kewley:2013ab,Maier:2014aa} and abundance scale \citep[e.g.,][]{Steidel:2014aa,Nicholls:2017aa}, can significantly affect the metallicity estimations. These parameters have not been well studied, particularly at high-redshift. Our local analogs of high-redshift galaxies providegreat opportunities to study what constrains these key input parameters in photoionization models (F. Bian et al. in preparation).

\section{Conclusion}\label{conclusion}
In this paper, we investigate the relations between the direct $T_e$-based metallicity and strong metallicity diagnostic line ratios in the stacked spectra of local analogs and normal SDSS galaxies. These analogs are selected to share the same location of $z\sim2$ star-forming galaxies in the {\oiii/\hb} versus {\nii/\ha} BPT diagram. They closely resemble the interstellar medium properties, including high ionization parameters and electron densities, of $z\sim2$ star-forming galaxies. These analogs provide a great opportunity to improve our understanding of the empirical metallicity calibrations in high-redshift galaxies. We summarize the main results of the paper as follows:

1. We select a sample of local analogs of high-redshift galaxies from the SDSS survey, which is located on the $z\sim2.0$ star-formation sequence on the BPT diagram. The ionization parameters and electron densities in these analogs are comparable to those in $z\sim2$ star-forming galaxies. Moreover, these galaxies show the exact the same behaviors relative to the local star-forming galaxies in different diagnostic diagrams. We also select a sample reference galaxies for a comparison. 

2. We generate a series of stacked spectra for the sample of local analogs and SDSS reference galaxies in different $N2=\log$({\nii/\ha}) bins. With the high S/N {\oau} detected in the stacked spectra, we measure the oxygen abundance based on the direct $T_e$ method.

3. We establish relations between the direct $T_e$ oxygen abundance and the N2 ($\log$(\nii/h$\alpha$)) and O3N2 ($\log$((\oiii/\hb)/(\nii/\ha))) indicators for the local analogs and SDSS galaxies in the reference sample. The relations in the local analogs of high-redshifts do not follow those in the SDSS reference galaxies. This new empirical calibration is suitable to measure metallicity in high-redshift galaxies. 

4. We build up the relation between the direct $T_e$ oxygen abundance and other metallicity indicators, including  R23, O32,  $\log($\oiiid/\hb), and $\log$({\neiii}/{\oii}) indices, in our local analogs, which can be used to measure metallicity in high-redshift galaxies. 

5. We apply the new empirical calibrations based on our local analogs to a sample of star-forming galaxies at $z\sim2$. Our new N2 and O3N2 empirical calibrations minimize the systematic discrepancy between the N2-based and O3N2-based metallicities in the star-forming galaxies at $z\sim2$.  The N2 and O3N2 metallicities will be underestimated by 0.05-0.1 dex, if one simply applies the local metallicity calibration to high-redshift galaxies.

6. By comparing our results with the MAPPINGS photoionization models, we find that the ISM conditions, including ionization parameter and electron density, play important roles in metallicity measurements. 

\acknowledgments
We would like to thank the anonymous referee for providing constructive comments and help in improving the manuscript.  F.B. thanks C. Steidel, A. Shapley,  M. Pettini, X. Fan, and P. Martini for useful discussions of the work. L. K. gratefully acknowledges support from an ARC Laureate Fellowship (FL150100113).  M. D. gratefully acknowledges the support of the Australian Research Council (ARC) through Discovery project DP16010363.

Funding for the SDSS and SDSS-II has been provided by the Alfred P. Sloan Foundation, the Participating Institutions, the National Science Foundation, the U.S. Department of Energy, the National Aeronautics and Space Administration, the Japanese Monbukagakusho, the Max Planck Society, and the Higher Education Funding Council for England. The SDSS Web Site is http://www.sdss.org/.



\facility{SDSS}





\bibliography{paper}

\clearpage





\end{document}